\documentclass[twocolumn,preprintnumbers,amsmath,amssymb]{revtex4}
\usepackage{graphicx}% Include figure files
\usepackage{dcolumn}% Align table columns on decimal point
\usepackage{bm}% bold math

\def\jcp#1#2#3{{\it J. Chem. Phys.}, {\bf #1}, #2 (#3)}
\def\prl#1#2#3{{ Phys.  Rev.  Lett.} {\bf #1}, #2 (#3)}

\def\pla#1#2#3{Phys.  Lett.  A {\bf #1}, #2 (#3)}
\def\pre#1#2#3{Phys.  Rev.  E {\bf #1}, #2 (#3)}
\def\pra#1#2#3{Phys.  Rev.  A {\bf #1}, #2 (#3)}

\def\etl{$et~al.~$}
\def\beq{\begin{equation}}
\def\eqn{\end{equation}}
\begin{document}

\title{q--deformed logistic map with delay feedback}
\author{Manish Dev Shrimali and Subhashish Banerjee}
\affiliation{
Indian Institute of Technology Rajasthan, Jodhpur 342 011, India
}
\date{\today}
\begin{abstract}

The delay logistic map with two types of q--deformations: Tsallis and Quantum--group type are studied.
The stability of the map and its bifurcation scheme is analyzed as a function of the deformation and delay feedback parameters. Chaos is suppressed in a certain region of deformation and feedback parameter space.
The steady state obtained by delay feedback is maintained in one type of deformation while chaotic behavior
is recovered in another type with increasing delay.

\end{abstract}
\maketitle

\section{Introduction}

Theory of quantum integrable systems \cite{vd85,mj85} has initiated a new type of symmetry and associated with it mathematical objects called quantum groups. These are related to the usual Lie groups as quantum mechanics is related to its classical
limit. Physically a group quantization can lead to a kind of deformation of the group manifold, related to a physical (classical or quantum) system. q--deformation  of  many  classical Lie groups has stimulated much activity  in  the  pursuit  of  understanding  its  physical  meaning, due to the emergence of quantum group like features in many physical systems.
It has been realized that q--deformation effectively takes  into account  the interactions in  physical systems
\cite{ev92,cy91,cgy,rp93}. The q--deformation  is non-trivial  in the sense that the emerging deformed algebra is no longer linear. It would be constructive to study q--deformation in the context of dynamical systems.

One of the popular model of discrete nonlinear dynamical systems is logistic map \cite{cd}. The study of dynamical system with delay is important when the information is feedback with a measurable delay, e.g., due to the spatial extensiveness of the system and the finite velocity of propagation of information, or when the characteristic timescale of the system is smaller than the delay time \cite{r3}. The delayed equations have been used for modeling purposes in optics \cite{d1, d2}, chemistry \cite{d3}, and biological systems \cite{d4, d5}.
The delay feedback also plays an important role in controlling chaos \cite{ogy, pyragas}. The interplay of delay and nonlinearity plays a central role in self-organization and complex phenomena of dynamical systems and chaos is suppressed or controlled by stabilizing unstable periodic orbits with delay feedback \cite{r2, r6, r1, r5, rd}. In  another  work \cite{ar99},  the  normal logistic  and exponential  maps were  used to  study the  transition from  chaotic to regular dynamics induced by stochastic driving.

A one-dimensional logistic map is a non-linear difference equation
 \begin{equation}
x_{n+1} = \alpha x_n (1 - x_n), \label{logistic}
\end{equation}
where $\alpha$ is  a constant, and is taken to be positive in the rest of the paper. Also, $x_n$ denotes the value  of $x$ after $n$ iterations. Eq.~(\ref{logistic}) arises, for  e.g., in the  case of modeling of population growth  $\frac{d N(t)}{d t}  = r(t) N(t)$, where  $N(t)$ is the  population at a  time $t$  and $r(t)$  is the  difference between birth and death rates per head of the population.

As q--deformation essentially involves modification of a function such that in the limit of $q \rightarrow 1$ the usual function is obtained, there is no unique q-deformation for a function. On the other hand, delay transforms the dynamical state of the system. It is therefore natural to use q-deformations suitably in the study of non-linear systems with delay feedback. Here we discuss two forms of q--deformations of the logistic map, studied in the literature, but with the additional proviso that the map has a memory inbuilt into it, in the form of a feedback mechanism. This has the advantage of studying the system from a more realistic perspective as well as the possibility of having a chaos suppression mechanism inbuilt into the system.

The paper is setup as follows. In Section II, we discuss the basic deformations,  that will be studied, along with delay. The logistic equation deformed by the two prescriptions and undergoing a delayed feedback are studied in Sections III (A) and (B), respectively. Along with a numerical study of the interplay between the various parameters, a stability study of the two systems is also made. Section IV concludes the paper.

\begin{figure}[t]
\begin{center}
\includegraphics[width=0.4\textwidth]{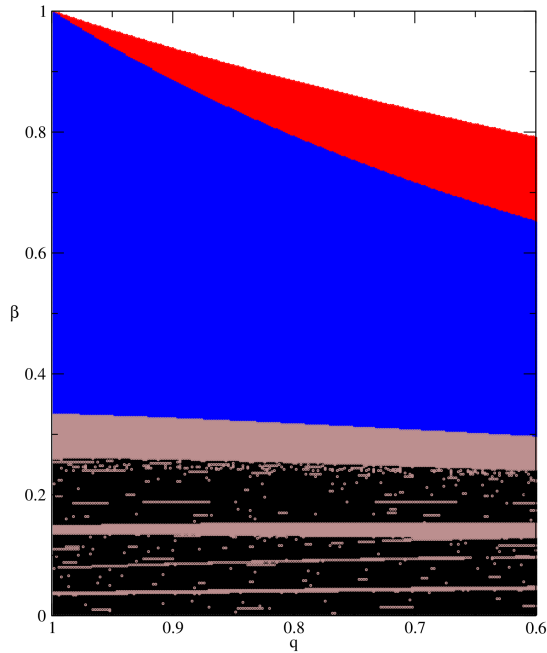}
\caption{The parameter space ($\epsilon$, $\beta$) showing chaotic (black color), periodic (brown color for period $>$ 2 and
blue for period 2), period 1
(non-zero steady state in blue color) and period 1 (zero steady state in white).
Here  nonlinearity parameter $\alpha = 4$,
delay parameter $\tau = 1$. As seen from the figure, with increase in coupling ($\beta$), for fixed q-deformation $\epsilon$, the system undergoes a transition from
the chaotic to the periodic regime, thereby highlighting the role played by the feedback on chaos suppression.}
\label{fig:fig1}
\end{center}
\end{figure}

\begin{figure}[t]
\begin{center}
\includegraphics[width=0.4\textwidth]{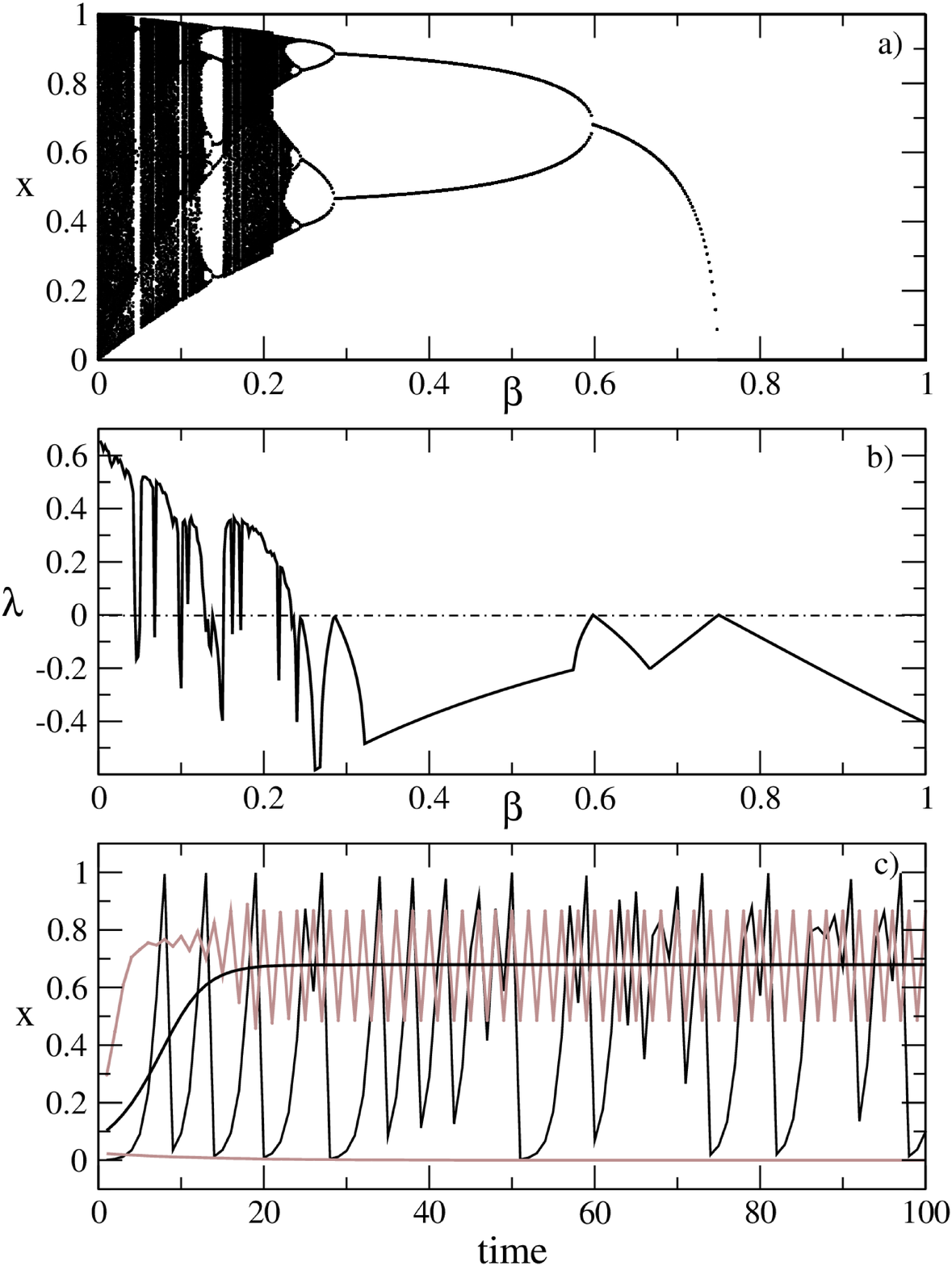}
\caption{a) Bifurcation diagram depicting $x$  with respect to $\beta$ for  $\alpha = 4$,
$\tau = 1$.  As seen from the figure, with increase in coupling ($\beta$), for fixed q-deformation $\epsilon = 0.5$, the system undergoes a transition from
the chaotic to the periodic regime, ultimately to the steady state, confirming the scenario depicted in Fig. (\ref{fig:fig2})
b), where the Lyapunov exponent $\lambda$  is plotted with respect to $\beta$ for  $\alpha = 4$,
$\tau = 1$. As seen from the figure, with increase in coupling ($\beta$), for fixed q-deformation $\epsilon$, the
system undergoes a transition from
the chaotic ($\lambda > 0$) to the periodic  ($\lambda < 0$) regime, ultimately to the steady  state.
c) Time series showing chaotic, period-2 and steady states for $\beta=0.0$, $0.5$, $0.7$, and $0.8$ respectively.}
\label{fig:fig2}
\end{center}
\end{figure}

\begin{figure}[t]
\begin{center}
\includegraphics[width=0.4\textwidth]{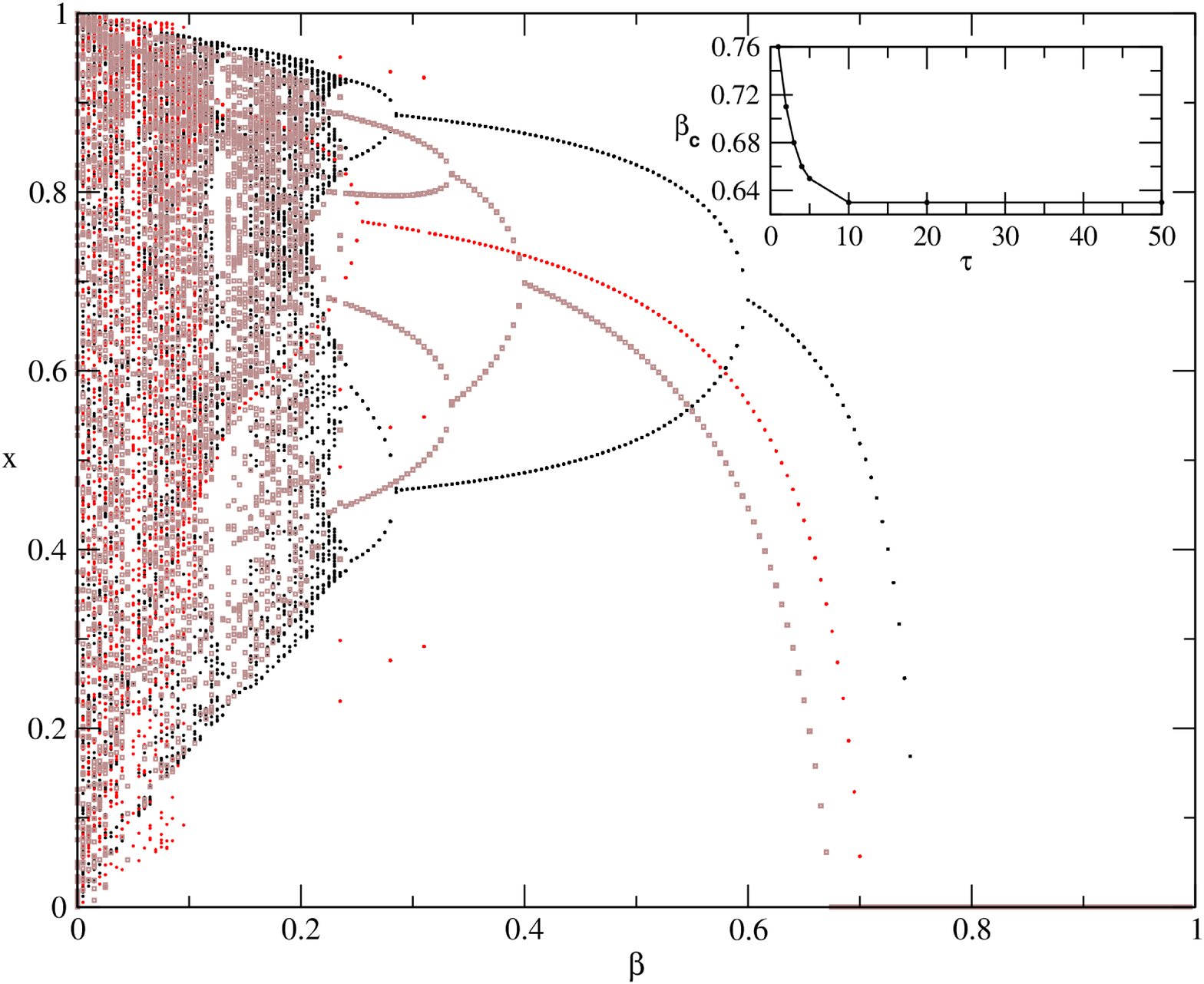}
\caption
{Bifurcation diagram depicting $x$  with respect to $\beta$ for  $\alpha = 4$ $q=0.5$,
$\tau = 1, 2$ and $3$.  System behavior is period two cycle, fixed point and chaotic for higher delay feedback strength
for $\tau=1, 2$ and $3$ respectively. In the inset, is plotted the critical value of the feedback parameter $\beta_c$,
for which the system goes into the steady state $x^{\star} = 0$.}
\label{fig:fig3}
\end{center}
\end{figure}

\section{$q$--deformation with delay}

A deformation of the logistic map (Eq.~\ref{logistic}), based on the non-extensive statistics of Tsallis \cite{tsallis} was proposed \cite{rs05}, in which the map
\begin{equation}
x_{n+1} = \alpha [x_n](1 - [x_n]), \label{qjagan}
\end{equation}
was considered.  Here $[x]  = \frac{x}{1+ (1-q) (1-x)}$, and $- \infty < q  < 2$  for $x$  in the  interval $[0,1]$.   An important difference  between (\ref{logistic})  and (\ref{qjagan})  is  that the deformed map (\ref{qjagan}) is concave in parts of $x$-space while the map without deformation (\ref{logistic}) is always convex.  Further, the use of (\ref{qjagan})  showed  the rare  phenomena of  the  co-existence  of attractors, i.e., the co-existence  of normal and chaotic behavior.

The logistic map with delay ($\tau$) feedback is given by:
\beq
x_{n+1} = f(x_n) = \alpha x_n (1-x_n) (1-\beta) + \beta x_{n-\tau} \label{logdelay}
\eqn
where, $\beta$ is the feedback amplitude and $\tau$ is the delay time. The analogous q-deformed map with delay is:
\begin{equation}
x_{n+1} = \alpha [x_n](1 - [x_n])(1-\beta) + \beta [x_{n-\tau}]. \label{qlogdelay}
\end{equation}

Another Quantum--group (Qu-group) type of q-deformation, of the logistic map, was proposed \cite{r02} as:
\begin{equation}
[x_{n+1}] = \alpha [x_n](1 - [x_n]), \label{qpartha}
\end{equation}
where
\begin{equation}
[x]  = \frac{1 - q^x}{1  - q}. \label{deformation}
\end{equation}
Here $q$ is real and $x$  is in the interval $[0,1]$.  This q-deformed logistic map is different from Eq.~(\ref{qjagan}). It is not possible to transform the q-deformed map, introduced in Eq. (\ref{qpartha}), to that in Eq.~(\ref{qjagan}). In particular, it is not possible to relate the $q$ parameter in Eq.~(\ref{qjagan}) to the q parameter in Eq.~(\ref{qpartha}). Further, in the proposed q--deformed map Eq.~(\ref{qpartha}), the left hand side is also q-deformed, in contrast to Eq.~(\ref{qjagan}). Thus in the space of q--deformed variables the q-deformed logistic map Eq.~(\ref{qpartha}), is the usual map. In mapping to ordinary space,
all the corresponding physical features emerge. This is not possible in the map Eq.~(\ref{qjagan}). In the limit $q \rightarrow 1$,  it is  seen that  $[x]  \rightarrow x$  and we  obtain the  usual logistic map (\ref{logistic}).

The corresponding deformed map with delay would be:
\begin{equation}
[x_{n+1}] = \alpha [x_n](1 - [x_n])(1-\beta) + \beta [x_{n-\tau}]. \label{qdelaypartha}
\end{equation}

What do we expect from such a study?  q-deformations simulate correlations in the system while the delayed feedback
brings in memory. An interplay of these two effects should help in understanding the mechanism of suppression of chaos in
systems that have inbuilt correlations.

\section{Analysis of the $q$--deformed logistic map with delay}

Here we take up the logistic map with delayed feedback and q-deformed according to both the prescriptions, i.e. according to Eqs. (\ref{qjagan}) (Tsallis type) and (\ref{qpartha}) (Qu-group type), respectively.

\subsection{Tsallis type of deformation}

\subsubsection{Stability Analysis: Analytical Results}

We make an analytical study of the effect of memory in Eq.~(\ref{qlogdelay}). This involves expanding the original equation
to a set of $\tau + 1$ equations, $\tau$ being the delay \cite{r3}. In order to capture the essence of
our q-deformed delayed logistic map, we take up the case of $\tau = 1$.  For the one-cycle stability, using
Eq.~(\ref{qlogdelay}):

\begin{center}
\begin{eqnarray}
& x_{n+1} =& \alpha (1-\beta)(2 - q)\frac{x_n(1-x_n)}{[1+(1-q)(1-x_n)]^2} \nonumber\\
&& + \beta \frac{y_n}{1+(1-q)(1-y_n)}, \nonumber\\
& y_{n+1} =& x_n.
\end{eqnarray}
\end{center}

The corresponding Jacobian matrix takes the form:

\begin{equation}
 J (x,y) = \left(\begin{array}{cc}
\alpha (1-\beta)(2 - q) \frac{2 - q - (3 - q)x_n}{[1+(1-q)(1-x_n)]^3} &
\frac{\beta (2-q)}{[1+(1-q)(1-y_n)]^2}\\
1 & 0
\end{array}\right),
\end{equation}
which for the trivial fixed point: $x = y = 0$ gives:
\begin{equation}
 J(0,0) = \left(\begin{array}{cc}
\frac{\alpha (1-\beta)}{(2-q)} & \frac{\beta}{(2-q)}\\
1 & 0
\end{array}\right). \label{jlogistic}
\end{equation}
From the characteristic equation of Eq. (\ref{jlogistic}), its eigenvalues are:
\begin{equation}
 \lambda_{1,2} = \frac{\alpha (1-\beta)}{2(2-q)} \pm
\frac{1}{2(2-q)} \sqrt{\alpha^2 (1-\beta)^2 + 4\beta (2-q)}. \label{eigenlog}
\end{equation}
From the above eigenvalues, the condition for the stability of the fixed point is obtained as:
\begin{equation}
 \beta > \frac{\alpha - (2-q)}{\alpha - 1}. \label{stablelog}
\end{equation}
These results are borne out by the numerical results shown below, where along with one-cycle stability, multi-cycle stability
is also analyzed.

\subsubsection{Numerical results}

We fixed the nonlinearity parameter $\alpha=4$ and delay time $\tau=1$ to study the effect of q--deformation and delay feedback in $(q, \beta)$ parameter space as shown in Fig.~\ref{fig:fig1}. For lower feedback strength $\beta$, system is in chaotic (in black) or in higher period (in brown) state. With sufficiently large delay feedback, system goes to period two cycle (in blue) and then to steady state $x^* \ne 0$ (in red) and then to $x^*=0$ (in white). The bifurcation diagram and Lyapunov exponent as a function of feedback strength $\beta$ is plotted in Figs.~\ref{fig:fig2}(b) and (c), respectively, for fixed q-deformation
$q=0.5$ which confirms chaos suppression via reverse period-doubling bifurcation. Time series in different
dynamical states, i.e. chaotic, period two cycle, and steady states are shown in Fig.~\ref{fig:fig2}(c).
When the delay time $\tau$ is further increased, the transition from chaos to steady state is seen to occur at lower values
of feedback strength $\beta$ and the critical value of feedback strength $\beta_c$, when the system approaches the steady state $x^*=0$, is $0.76$ for $\tau=1$ and approaches to $0.63$ with increasing $\tau$ (see Fig.~\ref{fig:fig3}).

Thus, in the Tsallis-type of q-deformed logistic map with delay, with increase in feedback, stability is seen for all delays;
in contrast to the situation in the corresponding map without deformation, where stability was observed for only odd
delays \cite{r3}.

\begin{figure}[t]
\begin{center}
\includegraphics[width=0.4\textwidth]{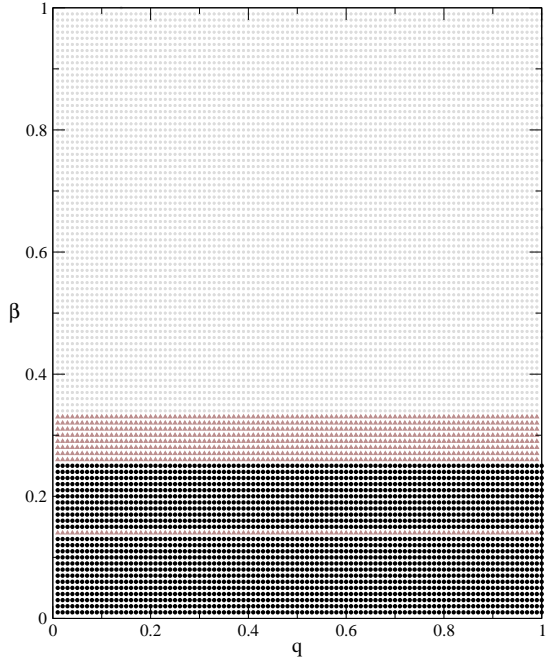}
\caption{The parameter space ($q$, $\beta$) showing chaotic (black color) and periodic (brown for period $>$ 2 and gray for
period 2). Here  nonlinearity parameter $\alpha = 4$, delay parameter $\tau = 1$. As seen from the figure, with increase in coupling ($\beta$), for any given q-deformation $\epsilon$, the system undergoes a transition from
the chaotic to the periodic--2 cycle regime, thereby highlighting the role played by the feedback on chaos suppression.}
\label{fig:fig4}
\end{center}
\end{figure}

\begin{figure}[t]
\begin{center}
\includegraphics[width=0.4\textwidth]{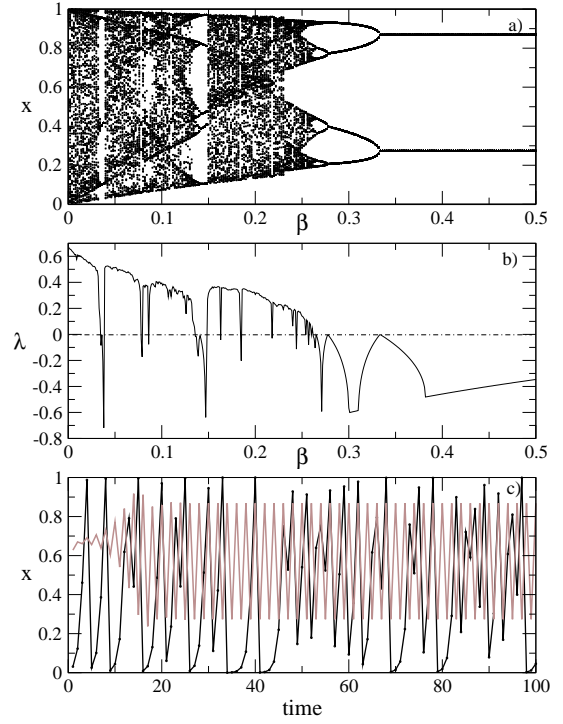}
\caption{a) Bifurcation diagram depicting $x$  with respect to $\beta$ for  $\alpha = 4$,
$\tau = 1$.  As seen from the figure, with increase in coupling ($\beta$), for fixed q-deformation $q = 0.5$, the system undergoes a transition from
the chaotic to the periodic--2 cycle regime confirming the scenario depicted in Fig. (\ref{fig:fig4}).
b) Lyapunov exponent $\lambda$  with respect to $\beta$ for  $\alpha = 4$,
$\tau = 1$. As seen from the figure, with increase in coupling ($\beta$), for fixed q-deformation $q$, the system undergoes a transition from the chaotic ($\lambda > 0$) to the periodic  ($\lambda < 0$) regime, ultimately to the period--2 cycle state.
c) Time series showing chaotic and period-2 states for $\beta=0.0$ and $0.4$ respectively.}
\label{fig:fig5}
\end{center}
\end{figure}

\begin{figure}[t]
\begin{center}
\includegraphics[width=0.4\textwidth]{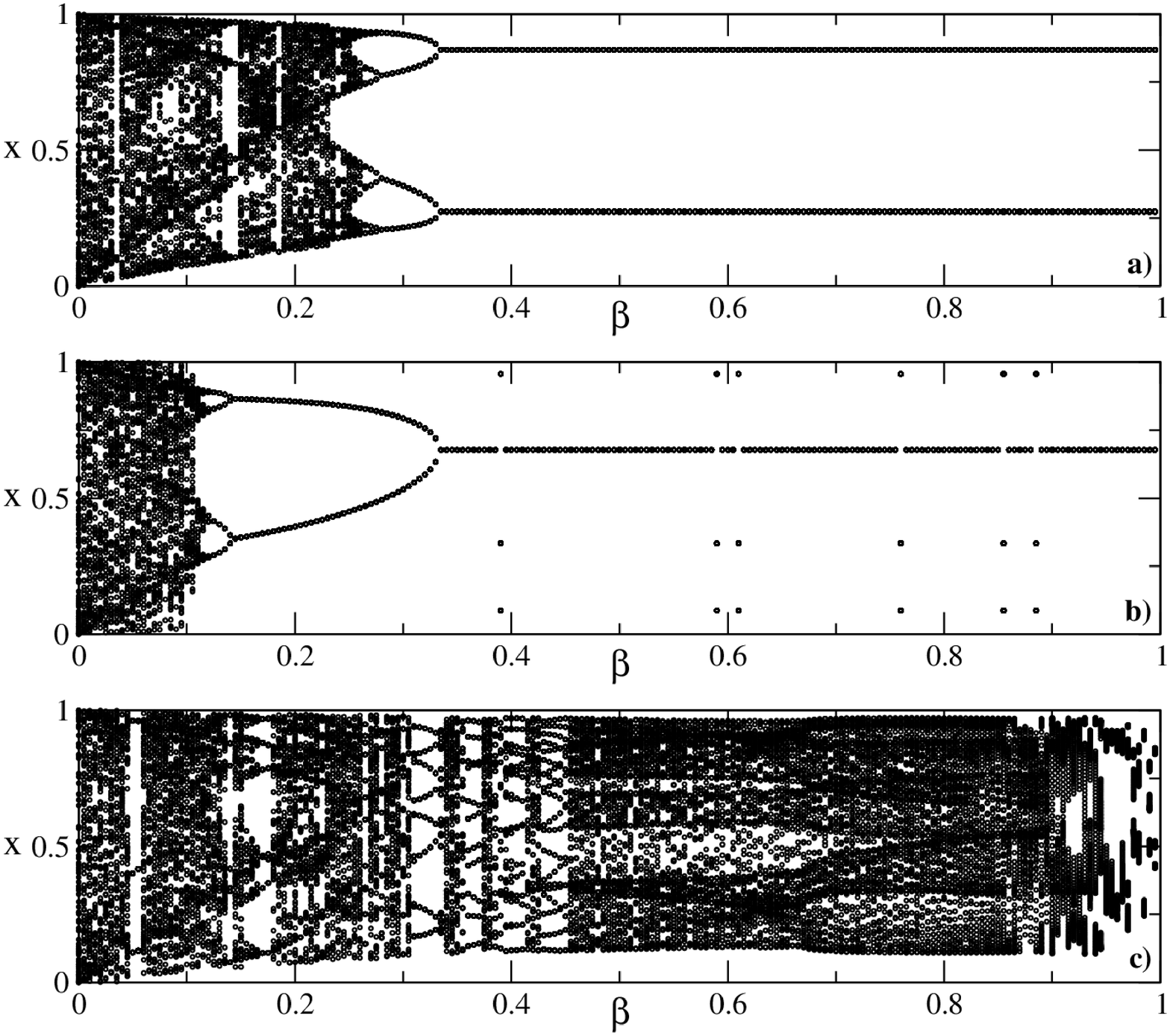}
\caption{Bifurcation diagram depicting $x$  with respect to $\beta$ for $\alpha = 4$ and $q=0.5$, for increasing delay time: $\tau = 1, 2$ and $3$, respectively. a) Period--2 cycle for $\tau=1$, b) fixed point for $\tau=2$, and c) chaotic behavior for $\tau=3$ are observed.
}
\label{fig:fig6}
\end{center}
\end{figure}

\subsection{Quantum--group type of deformation}

\subsubsection{Stability Analysis: Analytical Results}

We make an analytical study of the effect of memory in Eq. (\ref{qdelaypartha}). As before, in order to capture the essence of
our q-deformed delayed logistic map, we take up the case of $\tau = 1$.  For the one-cycle stability, using
Eq. (\ref{qdelaypartha}), we get the following two coupled equations:
\begin{eqnarray}
 X_{n+1} &=& 1 - \frac{\alpha (1 - \beta)}{(1-q)} (1-X_n)(X_n - q) - \beta (1-Y_n) \nonumber\\
Y_{n+1} &=& X_n,
\end{eqnarray}
where, for convenience in calculations, we have made the change in  variable : $q^{x_n} = X_n$.
The corresponding Jacobian matrix takes the form:
\begin{equation}
 J (X, Y) = \left(\begin{array}{cc}
\frac{\alpha (1-\beta)}{(1- q)} [2X_n - (q + 1)] & \beta \\
1 & 0
\end{array}\right),
\end{equation}
which for the trivial fixed point: $x = y = 0$, corresponding in the new variables to: $X = Y = 1$, gives:
\begin{equation}
 J(0,0) = \left(\begin{array}{cc}
\alpha (1-\beta) & \beta\\
1 & 0
\end{array}\right). \label{jqgroup}
\end{equation}
From the characteristic equation of Eq. (\ref{jqgroup}), its eigenvalues are:
\begin{equation}
 \lambda_{1,2} = \frac{\alpha (1-\beta)}{2} \pm
\frac{1}{2}\sqrt{\alpha^2 (1-\beta)^2 + 4\beta}. \label{eigenqdeform}
\end{equation}
From the above eigenvalues, the condition for the stability of the fixed point is obtained as:
\begin{equation}
 \beta < 1. \label{stableqgroup}
\end{equation}
Since these calculations are made in the new variable $X_n$, which is related to the original variable $x_n$ by
$x_n = \frac{\ln{X_n}}{\ln q}$; the result of Eq. (\ref{stableqgroup}) when interpreted in the original variable
implies that $x^* = 0$ will never stabilize under positive feedback.
This is  borne out by the numerical results shown below, where the one-cycle as well as multi-cycle stability
is  analyzed.

\subsubsection{Numerical results}

Again, we fixed the nonlinearity parameter $\alpha=4$ and delay time $\tau=1$ to study the effect of q--deformation and delay
feedback in $(q, \beta)$ parameter space, as shown in Fig.~\ref{fig:fig4}. For lower feedback strength $\beta$, system is in
chaotic (in black) or in higher period (in brown) state. With sufficiently large delay feedback, system goes to period
two cycle (in gray). The bifurcation diagram and Lyapunov exponent as a function of feedback strength $\beta$ is plotted in
Figs.~\ref{fig:fig5}(a) and (b), respectively for fixed q-deformation $q=0.5$ which confirms chaos suppression via reverse
period-doubling bifurcation to period two--cycle. Time series in different dynamical states, i.e., chaotic
and period two cycle are shown in Fig.~\ref{fig:fig5}(c). For delay time $\tau=2$, system goes to steady state with
sufficient feedback strength and with further increase of delay time system goes back to chaotic state as shown in
Fig.~\ref{fig:fig6}. For $\tau=2$, within period--1 cycle, some values of feedback strength $\beta$ take the system to period--3 cycle as shown in Fig.~\ref{fig:fig6}(b). It shows that chaos cannot be suppressed in Qu--group type of deformation for delay feedback with $\tau > 2$. We have studied the asymmetric Qu--group type of deformation case ($x_{n+1}=f([x_{n}])$) also and find similar results (not shown here).
Thus here the non-linearity is predominantly due to the form of the deformation chosen. In contrast to Tsallis type of
deformation, the Qu-group type of deformation, suppress  chaos to two-cycle as compared to
one--cycle.

\section{Conclusions}

In this work, we have studied the logistic map from the perspective of q--deformation and delay feedback. This enables us to study the interplay between the competitive features, viz. complexity, brought by the q--deformation and order, by delay feedback. Chaos is suppressed with feedback for both kinds of deformations, i.e., Tsallis and Qu--group type. However, for the Tsallis type of deformation, a steady state is achieved, an observation which validates its use in statistical mechanics of complex systems, where one would expect a system to eventually go to a steady state, while for the Qu-group type of deformation the period--2 cycle gets stabilized. With increasing delay time, the transition to steady state is obtained at lower values of feedback strength in Tsallis type of q--deformation while chaotic state is recovered in Qu--group type of q--deformation with increasing delay time.

%\vskip -.5cm

%\newpage

\end{document}